\documentclass[12pt]{article}
\usepackage{amsmath,amssymb}

\newcommand{\bea}{\begin{eqnarray}}
\newcommand{\eea}{\end{eqnarray}}
\newcommand{\be}{\begin{equation}}
\newcommand{\ee}{\end{equation}}
\newcommand{\vs}[1]{\vspace{#1 mm}}

\newcommand{\dsl}{\pa \kern-0.5em /}

\newcommand{\pa}{\partial}

\newcommand{\nn}{\nonumber\\}

\begin{document}
\topmargin 0mm
\oddsidemargin 0mm

\begin{flushright}

USTC-ICTS/PCFT-23-31\\

\end{flushright}

\vspace{2mm}

\begin{center}

{\Large \bf The  open string pair production, its enhancement and the physics behind}

\vs{10}

{\large J. X. Lu}

\vspace{4mm}

{\em
Interdisciplinary Center for Theoretical Study\\
 University of Science and Technology of China, Hefei, Anhui
 230026, China\\
 \medskip
 Peng Huanwu Center for Fundamental Theory, Hefei, Anhui 230026, China\\ 
 %\vs{4}
}

\end{center}

\vs{10}

\begin{abstract}
 Why adding a collinear magnetic field to the electric one on a D3 brane in a system of two parallel separated D3 branes can enhance the open string pair production?  How is the open string pair production rate related to the QED ones in the weak field limit?  We answer these questions and report the somehow expected but still remarkable relation between the  weak field non-perturbative stringy rate and the corresponding QED ones in this letter.   
\end{abstract}

\newpage

The famous Schwinger pair production \cite{Schwinger:1951nm} is a physically important phenomenon of quantum electrodynamics (QED) due to the vacuum instability in the presence of an electric field.  It is natural to ask if there exists an analogous process in various string theories.  This was pursued in unoriented bosonic string and Type I superstring in  \cite{Bachas:1992bh, Porrati:1993qd}. The present author and his collaborators made a systematic study of this process  in the oriented Type II superstring theories in
%\cite{Lu:2009yx, Lu:2009au, Lu:2009pe, Lu:2017tnm, Lu:2018suj, Lu:2018nsc, Lu:2019ynq, Jia:2018mlr, Jia:2019hbr, Lu:2020hml, Lu:2023jxe}  
\cite{Lu:2009yx} - \cite{Lu:2023jxe} by focusing on the non-perturbative 1/2 BPS D branes of these theories.

The  quantum fluctuations of  an isolated Dp brane  in the weak string coupling can be described by open strings with the two ends of each string, carrying equal but opposite U(1) charge,  attached on the D brane. The virtual two charged ends appear to a brane observer as a virtual charged/anti-charged pair and as such one is wondering if there is a certain probability to produce such a pair once a constant worldvolume electric field is applied. The answer is simply no due to that unlike in QED, the two ends are connected by an open string. The applied electric field can only stretch the open string but cannot break it since the applied field needs to be below its critical one to maintain the D brane itself largely stable. In other words,  an isolated Dp brane cannot give rise to the open string pair production in Type II string theories.  Further discussion of this from various angles has been given in \cite{Lu:2023jxe}.  This result appears in sharp contrast with the usual Schwinger pair production \cite{Schwinger:1951nm} for  the case of $p = 3$, the one mimicking our own world.

The possible simplest case for having such a  pair production while keeping the brane system largely stable is to consider a system of two Dp branes with $p > 0$, placed parallel at a separation. In addition to the open string fluctuations  discussed above, there is now a new one consisting of a pair of a virtual  open string and a virtual anti open string connecting the two D branes. The two ends of the pair on one Dp brane carry equal but opposite U(1) charge associated with this brane while the other two ends of the pair  on the other Dp carry also the equal but opposite U(1) charge associated with the other brane.  If an electric field is applied on either of the two D branes, there is indeed a certain probability to separate the virtual open string from the anti open string and make the virtual pair become real \cite{Lu:2017tnm, Lu:2018suj, Lu:2018nsc}. This is just like the usual Schwinger pair production to either brane observer.  Moreover the produced pair originates from the virtual pair connecting the two Dp branes, i.e. along the directions transverse to the branes, therefore a detection of this pair production can provide a means to reveal the existence of extra dimensions \cite{Lu:2018nsc}.

 We have shown that  adding a magnetic flux in a specific way with respective to the electric one will enhance most effectively the pair production \cite{Lu:2017tnm}. At a first look, this is sort of counterintuitive since a pure magnetic flux cannot produce the pair production. We will explain this enhancement and provide the physics behind in this Letter. 
 
 It has been shown \cite{Lu:2017tnm, Lu:2018suj} that given the magnitudes of the respective applied electric flux and magnetic flux,  the enhancement is largest when the two do not share any common field strength index. This implies that we need the spatial dimensionality of the Dp branes to be $p \ge 3$.  We also showed in \cite{Jia:2018mlr} that adding any additional magnetic flux does not help in this regard. In practice, the largest rate occurs for $p = 3$ \cite{Lu:2017tnm, Lu:2018nsc}, the braneworld mimicking our own world, when  the electric flux and the magnetic one remain the same for all $p \ge 3$ cases.
 
So considering the D3/D3 system is not only for concreteness but also has its own significance. Without further ado, we focus in the following on the system of two D3 branes with each brane carrying the following flux
 \be\label{eg3case}
\hat F^{a}=\left( \begin{array}{cccc}
0 &{\hat f}_{a}& 0 &0 \\
-{\hat  f}_{a}&0&0&0\\
0&0&0& {\hat g}_{a}\\
0&0& - {\hat g}_{a}&0\\
 \end{array}\right),
 \ee
 where $a = 1, 2$ and $\hat F_{\alpha\beta} = 2\pi \alpha' F_{\alpha\beta}$ with $\alpha, \beta = 0, 1, 2, 3$. Here $F_{\alpha\beta}$ is the usual field strength defined on the brane and $2\pi \alpha'$ is the inverse of the fundamental string tension, giving a dimensionless field strength $\hat F_{\alpha\beta}$. Simply from the corresponding Born-Infeld factor, one can deduce $|{\hat f}_{a}| < 1$ and $|{\hat g}_{a}| < \infty$ to maintain the brane stable. 
 
 The closed string tree-level cylinder interaction amplitude between the two D3 can be computed using the D brane boundary state representation  \cite{Di Vecchia:1999fx}, following \cite{Lu:2018suj,Jia:2019hbr}. The corresponding open string one-loop annulus amplitude can be obtained by applying a Jacobi transformation to the cylinder one and is given as
 \be \label{annulusI}
\Gamma =   \frac{4  V_4 |\hat f_1 - \hat f_2| |\hat g_1 - \hat g_2|}{(8 \pi^2 \alpha')^2} \int_0^\infty \frac{d t}{t} e^{- \frac{y^2 t}{2\pi \alpha'}} \frac{(\cosh \pi \nu'_0 t - \cos\pi \nu_0 t)^2}{\sin\pi \nu_0 t \,\sinh\pi \nu'_0 t} \prod_{n = 1}^\infty Z_n,
\ee
where $y$ is the brane separation and 
\be
Z_n = \frac{\prod_{j =1}^2 [1 - 2 \,e^{(-)^j \pi \nu'_0 t} |z|^{2n} \cos\pi \nu_0 t + e^{(-)^j 2\pi \nu'_0 t} |z|^{4n}]^2}{(1 - |z|^{2n})^4 (1 - 2\, |z|^{2n} \cos2\pi\nu_0 t + |z|^{4n}) \prod_{j = 1}^2(1 - e^{(-)^{(j - 1)}2\pi\nu'_0 t} |z|^{2n})},
\ee
with $|z| = e^{- \pi t}$.  In the above, the parameters $\nu_{0} \in [0, \infty)$ and $\nu'_{0} \in [0, 1)$ are  determined via
\be\label{egparameter}
\tanh\pi\nu_{0} = \frac{|\hat f_{1} - \hat f_{2}|}{1 - \hat f_{1} \hat f_{2}}, \quad \tan \pi \nu'_{0} = \frac{|\hat g_{1} - \hat g_{2}|}{1 + \hat g_{1}\hat g_{2}}.
\ee
Note that $\nu_{0} \to \infty$ when either $|{\hat f}_{a}| \to 1$, the critical value of electric flux.  For large $t$, the integrand of the above amplitude behaves like $e^{- \frac{t}{2\pi\alpha'} (y^{2} - 2 \pi^{2} \alpha' \nu'_{0})}/t$ which vanishes when $y \ge \pi \sqrt{2 \nu_{0}' \alpha'}$ but blows up when $y <   \pi \sqrt{2 \nu_{0}' \alpha'}$, indicating a tachyonic instability at this separation. 

The factor $\sin\pi\nu_{0} t$ in the denominator of the integrand gives an infinite number of simple poles of this integrand, occurring at $t_{k} = k/\nu_{0}$ with $k = 1, 2, \cdots$ along the positive t-axis, signaling the decay of this system via the open string pair production at these poles. The non-perturbative decay rate of this system can be computed following \cite{Bachas:1992bh} as the residue of the integrand at these poles times $\pi$ and is given as
\be\label{decayrate}
 {\cal W} = - \frac{2 \,{\rm Im} \Gamma}{V_4} = \frac{8   |{\hat f}_1 - {\hat f}_2||{\hat g}_1 - {\hat g}_2|}{(8\pi^2 \alpha')^2} \sum_{k = 1}^\infty (-)^{k - 1} \frac{\left[\cosh\frac{\pi k \nu'_0}{\nu_0} - (-)^k\right]^2}{k \sinh \frac{\pi k \nu'_0}{\nu_0}} \, e^{- \frac{ k y^2}{2\pi \alpha' \nu_0}} \prod_{n = 1}^\infty Z_{k, n} 
\ee
where 
\be\label{fn}
Z_{k, n}  = \frac{\left(1 -  (-)^k \, e^{- \frac{2 n k \pi}{\nu_0} (1 - \frac{\nu'_0}{2 n})}\right)^4 \left(1 - (-)^k  \, e^{- \frac{2 n k \pi}{\nu_0}(1 + \frac{\nu'_0}{2 n})}\right)^4}{\left(1 - e^{- \frac{2 n k \pi}{\nu_0}}\right)^6 \left(1 -   \, e^{- \frac{2 n k \pi}{\nu_0} (1 - \nu'_0/ n)}\right) \left(1 -  \, e^{- \frac{2 n k \pi}{\nu_0}(1 + \nu'_0 / n)}\right)}.
\ee
The open string pair production rate is just the $k = 1$ term of the above decay rate, following \cite{nikishov} (see the recent discussion in \cite{Lu:2023jxe}), and is
  \be \label{pprate} {\cal W}^{(1)} = 
\frac{8 \,|\hat f_1 - \hat f_2||\hat g_1 - \hat g_2|}{(8\pi^2\alpha')^{2} }  e^{- \frac{y^2}{2\pi \nu_0 \alpha'}}  \frac{\left[\cosh
\frac{\pi \nu'_0}{\nu_0} + 1 \right]^2}{ \sinh
\frac{\pi \nu'_0}{\nu_0}} Z_{1} (\nu_{0}, \nu'_{0}),
\ee
where
\be\label{z1}
Z_{1} (\nu_{0}, \nu'_{0}) =  \prod_{n = 1}^\infty
\frac{\left[1 + 2  e^{- \frac{2 n  \pi}{\nu_0}} \cosh
\frac{\pi \nu'_0}{\nu_0} + e^{- \frac{4 n 
\pi}{\nu_0}}\right]^4}{\left[1 - e^{- \frac{2 n 
\pi}{\nu_0}}\right]^6 \left[1 - e^{- \frac{2  \pi}{\nu_0}(n -
\nu'_0)}\right]\left[1 - e^{- \frac{2  \pi}{\nu_0}(n +
\nu'_0)}\right]}.
\ee
As said at the outset, to keep the system largely stable, the electric flux applied on each brane is below its critical one (otherwise, the cascading pair production occurs which can seen from the above rates  when $\nu_{0} \to \infty$) and as shown in \cite{Lu:2020hml}, we have $\nu_{0} \ll 1$ in general.  With this, we have from (\ref{z1}) $Z_{1} \to 1$. In other words, the dominant contribution to the pair production comes from the lowest modes of the open string pairs connecting the two D3. Further in the weak magnetic flux limit, we have from (\ref{egparameter}) $\pi \nu'_{0} \approx |{\hat g}_{1} - {\hat g}_{2}|$ in addition to $\pi \nu_{0} \approx |{\hat f}_{1} - {\hat f}_{2}|$.  For simplicity, we take one D3 as a visible one such as ours carrying non-vanishing fluxes while the other as the hidden one carrying no fluxes and set the non-vanishing electric flux, say, ${\hat f}_{1} = 2\pi \alpha' e E$, and the magnetic one as ${\hat g}_{1} = 2\pi \alpha' e B$, with $E$ and $B$ the usual electric field and magnetic field, respectively. 

With all these, the open string pair production rate (\ref{pprate}) becomes a neat one as
 \be\label{eg3pprate-new}
{\cal W}^{(1)} = \frac{ 2 (e E) (e B)}{(2 \pi)^2 } \frac{\left[\cosh\frac{\pi B}{ E} +1\right]^2}{\sinh \frac{\pi B}{E}} \, e^{- \frac{  \pi m^{2} (y)}{e E }},
\ee
where we have introduced a mass scale,  which is the mass for the lowest modes of the open string connecting the two D3,  
\be\label{mscale}
m (y) =  T_{f} y = \frac{y}{2\pi \alpha'}.
\ee

Let us pause here briefly to understand the open string modes contributing to the above pair production rate.  In the absence of both $E$ and $B$, the mass spectrum for the open string connecting the two D3 is
 \be\label{mass-level}
\alpha' M^{2} = -\alpha' p^{2} =  \left\{\begin{array}{cc}
\frac{y^{2}}{4\pi^{2} \alpha'} + N_{\rm R} &  (\rm R-sector),\\
\frac{y^{2}}{4\pi^{2} \alpha'}  + N_{\rm NS}  - \frac{1}{2} & (\rm NS-sector),
\end{array}\right.
\ee
where $p = (k, 0)$ with $k$ the momentum along the brane worldvolume directions, $N_{\rm R}$ and $N_{\rm NS}$ are the standard number operators in the R-sector and NS-sector, respectively, as
\be\label{NP}
N_{\rm R} = \sum_{n = 1}^{\infty} (\alpha_{- n} \cdot \alpha_{n} + n d_{-n} \cdot d_{n}),\quad N_{\rm NS} = \sum_{n = 1}^{\infty} \alpha_{- n} \cdot \alpha_{n} + \sum_{r = 1/2}^{\infty} r d_{- r} \cdot d_{r}.
\ee
The R-sector gives fermions with $N_{\rm R} \ge 0$ while the NS-sector gives bosons with $N_{\rm NS} \ge 1/2$.  The $N_{\rm R} = 0$, $N_{\rm NS} = 1/2$ give the usual
massless $4 (8_{\rm F} + 8_{\rm B})$ degrees of freedom(DOF) or the 4D {\cal N} = 4  U(2) Super Yang-Mills (SYM) from the D3 worldvolume view,  when $y = 0$.  Among these, $2 (8_{\rm F} + 8_{\rm B})$  become massive ones, all with mass $ T_{f} y = y/(2\pi \alpha') $ due to unbroken SUSY, when $y\neq 0$. This just reflects  $U (2)  \to U(1) \times U(1)$ when $y = 0 \to y \neq 0$.  The two broken generators give 16 pairs of massive charged/anti-charged DOF with respect to either brane observer's own unbroken U(1), counting  5 scalar pairs, 4 spinor pairs and one vector pair. In other words,  we have 16 pairs  contributing to the pair production rate (\ref{eg3pprate-new}). 

The $B = 0$ pair production rate can be obtained from (\ref{eg3pprate-new}) by taking $B \to 0$ limit as
\be
{\cal W}^{(1)} (B = 0) = \frac{8 (e E)^{2}}{\pi (2 \pi)^{2}} e^{- \frac{\pi m^{2}}{e E}} =  \frac{16 (e E)^{2}}{ (2 \pi)^{3}} e^{- \frac{\pi m^{2}}{e E}}.
\ee
 The enhancement of the rate due to the presence of the magnetic flux $B$ is therefore
  \be\label{rate-enhancement}
\frac{{\cal W}^{(1)} (B \neq 0)}{{\cal W}^{(1)} (B = 0)}  = \frac{1}{4} \frac{ \pi  B}{E}  \frac{\left[\cosh\frac{\pi B}{ E} +1\right]^2}{\sinh \frac{\pi B}{E}} 
\ee
which is larger than unity when $B/E \sim {\cal O} (1)$ and becomes 
\be
\frac{1}{8} \frac{\pi B}{E} \, e^{\frac{\pi B}{E}} \gg 1,
\ee
when $B/E \gg 1$.  We now come to explain the physics behind such an enhancement.  For this, let us first make a comparison of the present rate (\ref{eg3pprate-new}), denoting now as ${\cal W}^{\rm{string}} = {\cal W}^{(1)}$, with the various QED rates for a charged scalar pair, a charged spinor pair and a charged vector pair, respectively.  For a massive charged scalar pair and a massive charged spinor pair, the corresponding rates \cite{nikishov, Kim:2003qp} are, respectively, as
 \be\label{scalar/spinorQED}
 {\cal W}^{\rm{QED}}_{\rm scalar} = \frac{(e E) (e B)}{2 (2\pi)^{2}} {\rm csch} \left(\frac{\pi B}{E}\right) \, e^{- \frac{\pi m_{0}^{2}}{e E}}, \quad
{\cal W}^{\rm{QED}}_{\rm spinor} = \frac{(e E) (e B)}{(2\pi)^{2}} \coth\left(\frac{\pi B}{E}\right) \, e^{- \frac{\pi m_{1/2}^{2}}{e E}}.
\ee 
For a massive charged vector pair, the QED rate \cite{Kruglov:2001cx} is
\be\label{w-bosonr}
{\cal W}^{\rm{QED}}_{\rm vector} = \frac{(e E) (e B)}{ 2 (2 \pi)^{2}} \frac{ 2 \cosh \frac{2 \pi B}{E} + 1}{ \sinh \frac{\pi B}{E}} \, e^{- \frac{\pi m^{2}_{1}}{e E}}.
\ee
In order to make a sensible comparison, we need to set all the masses for scalar, spinor and vector to be the same as the mass scale $m$ as in the rate (\ref{eg3pprate-new}). Let us first consider the case of $B = 0$ and we have
\bea\label{rate-ratio}
 {\cal W}^{\rm{QED}}_{\rm scalar} &=& \frac{ (e E)^{2}}{(2 \pi)^{3}}\, e^{- \frac{\pi m^{2}}{e E}}, \quad {\cal W}^{\rm{QED}}_{\rm spinor} = \frac{ 2 (e E)^{2}}{(2 \pi)^{3}} \, e^{- \frac{\pi m^{2}}{e E}},\quad
   {\cal W}^{\rm{QED}}_{\rm vector} = \frac{3 (e E)^{2}}{(2 \pi)^{3}}\, e^{- \frac{\pi m^{2}}{e E}}, \nn
   {\cal W}^{\rm{string}} &=& \frac{16 (e E)^{2}}{(2 \pi)^{3}}\, e^{- \frac{\pi m^{2}}{e E}},
\eea
With respective to the scalar pair, it is clear that the above rates are simply the counting, in each case, of the underlying DOF, and we have
\be
{\cal W}^{\rm{QED}}_{\rm vector} = 3 \,{\cal W}^{\rm QED}_{\rm scalar},\quad  {\cal W}^{\rm{QED}}_{\rm spinor} = 2\, {\cal W}^{\rm QED}_{\rm scalar},\quad
{\cal W}^{\rm{string}} = 16\,  {\cal W}^{\rm QED}_{\rm scalar}.
\ee
 The above clearly imply the following relation 
 \be\label{rate-relation}
{\cal W}^{\rm{string}}  = 5\, {\cal W}^{\rm{QED}}_{\rm scalar}  + 4\, {\cal W}^{\rm{QED}}_{\rm spinor}  + {\cal W}^{\rm{QED}}_{\rm vector}.
\ee 
 We would like to stress that the left side is from a non-perturbative stringy computation and needs the presence of the so-called hidden D3 while the right side is purely from the (1 + 3)-dimensional QED non-perturbative computations which say nothing  about the hidden D3 other than our own world.  The two approaches are completely different but the end result is the same, somehow expected but quite remarkable.
 
 Let us now consider the other limiting  case of $B \gg E$. From the rates (\ref{scalar/spinorQED}), (\ref{w-bosonr}) and (\ref{eg3pprate-new}),  we have
 \bea\label{largeE}
 {\cal W}^{\rm{QED}}_{\rm scalar} &=& \frac{(e E) (e B)}{ (2\pi)^{2}}\, e^{- \frac{\pi (m^{2} + e B)}{e E}},\quad  {\cal W}^{\rm{QED}}_{\rm spinor} = \frac{(e E) (e B)}{(2\pi)^{2}}  \, e^{- \frac{\pi m^{2}}{e E}},\nn
 {\cal W}^{\rm{QED}}_{\rm vector} &=& \frac{(e E) (e B)}{  (2 \pi)^{2}}  \, e^{- \frac{\pi (m^{2} - e B)}{e E}}, \quad {\cal W}^{\rm{string}} = \frac{  (e E) (e B)}{(2 \pi)^2 }  \, e^{- \frac{  \pi (m^{2} - e B)}{e E }}, 
\eea 
 where we have also set $m_{0} = m_{1/2} = m_{1} =  m$. The first thing we notice is that the prefactor for all the rates is the same as that for the scalar one. This  implies that when $B \gg E$, we have only one pair contributing to the underlying rate for each case. We will come to explain the physics behind  along with the rate enhancement in a moment. For now, let us examine directly the various rates in (\ref{largeE}). It is clear for $B \gg E$
\be
 \frac{{\cal W}^{\rm QED}_{\rm scalar}}{{\cal W}^{\rm{QED}}_{\rm vector}}  = e^{- \frac{2 \pi B}{ E}} \to 0,\quad  \frac{ {\cal W}^{\rm{QED}}_{\rm spinor}}{{\cal W}^{\rm{QED}}_{\rm vector}} = e^{- \frac{\pi B}{ E}} \to 0.
\ee 
and 
\be\label{largeEC}
{\cal W}^{\rm{string}} =  {\cal W}^{\rm{QED}}_{\rm vector} = \frac{(e E) (e B)}{  (2 \pi)^{2}}  \, e^{- \frac{\pi (m^{2} - e B)}{e E}} .
\ee 
In other words, for $B \gg E$, the relation (\ref{rate-relation}) continues to hold.  This makes one wonder if this relation holds in general, i.e., for general $E$ and $B$. One can check directly that this remains indeed true using the respective rates (\ref{eg3pprate-new}), (\ref{scalar/spinorQED}) and (\ref{w-bosonr}) when we set $m_{0} = m_{1/2} = m_{1} =m$, once again a remarkable but physically expected result. 

We now come to explain the rate behaviors in (\ref{largeE}) and the rate enhancement in the presence of a magnetic flux $B$ with $B/E > {\cal O} (1)$. For this, let us consider a charged particle with its spin $S$ and mass $m_{S}$ moving in a magnetic field  ${\bf B} = B\, {\bf e}_{x}$ (being consistent with out choice (\ref{eg3case})). The energy spectrum of this particle is
\be\label{chargedp}
E^{2}_{(S, S_{x})} = (2 N + 1) e B \mp g_{S} e {\bf B} \cdot {\bf S} + m^{2}_{S},
\ee
with $e$ the unit positive charge, $g_{S}$ the gyromagnetic ratio ($g_{S} = 2$) and $N$ the Landau level. Here $\mp$ correspond to the positive/negative charge carried by the particle, respectively. So for the lowest Landau level ($N = 0$), we have the following mass splittings for positively charged particle\\

\begin{tabular}{|c|c|c|c|}
\hline
S& 0 & 1/2 & 1\\
\hline
 & & $E^{2}_{(\frac{1}{2}, - \frac{1}{2})} = m^{2}_{\frac{1}{2}} + 2 e B$ & $E^{2}_{(1, - 1)} = m^{2}_{1} + 3 e B$\\
$E^{2}_{(S, S_{x})}$ & $E_{(0, 0)}^{2} = m^{2}_{0} + e B$ & & $E^{2}_{(1, 0)} = m^{2}_{1} + e B$ \\
 & & $E^{2}_{(\frac{1}{2}, \frac{1}{2})} = m^{2}_{\frac{1}{2}}$ & $E^{2}_{(1, 1)} = m^{2}_{1} - e B$\\
 \hline  
\end{tabular}\\
 
 \noindent 
It is clear from the above table that for a charged scalar, its effective mass increases;  for a positively charged spinor, the effective mass for the spin-up ($S_{x} > 0$) polarization  remains unchanged while the effective mass increases for the spin-down ($S_{x} < 0$)  polarization, and for the positively charged vector field, the effective masses for the spin-down and $S_{x} = 0$ polarizations both increase while for the spin-up polarization its effective mass reduces, which is the key for the enhancement of the open string pair production  in (\ref{eg3pprate-new}). For a  negatively charged spinor or vector particle, we need to switch the polarizations in the above table. For example, the spin-up case for the positively charged particle becomes the spin-down case for the negatively charged particle.  This is also necessary for the pair production due to the (total) net-zero  charge and angular momentum conservations for the pair, respectively. 

If we use the polarization $S_{x}$ of positively charge  mode to denote the pair, implying the polarization of the negatively charged mode in the pair being $- S_{x}$, we then have the pair production rate for each such pair for large $B/E$ as
\be
{\cal  W}^{\rm QED}_{(S, S_{x})} = \frac{  (e E) (e B)}{(2 \pi)^2 } e^{- \frac{\pi E^{2}_{(S, S_{x})}}{e E}},
\ee 
where $E_{(S, S_{x})}$ is the energy for each positively charged mode. With this, it is easy to check for large $B/E$, for $S = 1/2$, 
\be
\frac{{\cal W}^{\rm QED}_{(1/2, - 1/2)}}{{\cal W}^{\rm QED}_{(1/2, 1/2)}} = e^{- \frac{2\pi B}{E}} \to 0,
\ee
and for $S = 1$,
\be
\frac{{\cal W}^{\rm QED}_{(1, -1)}}{{\cal W}^{\rm QED}_{(1, 1)}} = e^{- \frac{4 \pi B}{E}} \to 0, \quad  \frac{{\cal W}^{\rm QED}_{(1, 0)}}{{\cal W}^{\rm QED}_{(1, 1)}} = e^{- \frac{2 \pi B}{E}} \to 0.
\ee 
So the above explains the results in (\ref{largeE}). In other words,  for large $B/E$, only the pair with the polarizations having the lowest energy contributes to the rate.  For the open string pair production rate 
(\ref{eg3pprate-new}), the enhancement for large $B/E$ comes from the pair of the $S_{x} =   1$ polarization of the positively charged vector mode along with the $S_{x} = - 1$ negatively charged one since their effective mass is actually lowered and this makes the virtual pair become a real one much easier. So we now understand the underlying physics for both the rate results in (\ref{largeE})  and the enhancement of the open string pair production for reasonable large $B/E$.

In summary,  the underlying physics to the enhancement of the stringy pair production is the lowering of the mass of the charged/anti-charged vector pair with the respective $S_{x} = 1$/$S_{x} = - 1$ polarizations  when a large collinear magnetic flux is added such that the production of this pair becomes much easier.  We derive for the first time the somehow expected but still remarkable relation between the weak field stringy rate and those rates from QED as given in (\ref{rate-relation}) which is valid for a general collinear $E$ and $B$.  The stringy rate needs the presence of a 
 hidden D3  while the QED rates say nothing about this.  This tells their difference of capability in probing the world beyond our visible one. Further the presence of the hidden D3 as required by the stringy approach can be a source of dark matter, for example. 

% put long equation here
%\end{widetext}

\section*{Acknowledgments}
The author acknowledges the support by grants from the NNSF of China with Grant No: 12275264 and 12247103.

\end{document}